\documentclass[prl,twocolumn,reprint,amsmath,amssymb,amsfonts,superscriptaddress,showpacs]{revtex4-1}

\usepackage{graphicx}
\usepackage{multirow}
\usepackage{color}
\usepackage{amsthm}
\usepackage{here}

\newcommand{\mb}[1]{\ensuremath{\mathbf{#1}}}

\newcommand{\mr}[1]{\ensuremath{\mathrm{#1}}}

\newcommand{\ovl}[2]{\ensuremath{\langle #1 | #2 \rangle}}

\newcommand{\bo} {\ensuremath{\mathcal{O}}}

\begin{document}

\title{Ultrafast Emission and Detection of a Single-Electron Gaussian Wave Packet:\\A Theoretical Study }
\author{Sungguen Ryu}
\affiliation{Department of Physics, Korea Advanced Institute of Science and Technology, Daejeon 34141, Korea}
\author{M. Kataoka}
\affiliation{National Physical Laboratory, Hampton Road, Teddington, Middlesex TW11 0LW, United Kingdom}
\author{H.-S. Sim}\email[]{hssim@kaist.ac.kr}
\affiliation{Department of Physics, Korea Advanced Institute of Science and Technology, Daejeon 34141, Korea}

\date{\today}

\begin{abstract} 
Generating and detecting a prescribed single-electron state is an important step towards solid-state fermion optics. We propose how to generate an electron in a Gaussian state, using a quantum-dot pump with gigahertz operation and realistic parameters. With the help of a strong magnetic field, the electron occupies a coherent state in the pump, insensitive to the details of nonadiabatic evolution. The state changes during the emission from the pump, governed by competition between the Landauer-Buttiker traversal time and the passage time. When the former is much shorter than the latter, the emitted state is a Gaussian wave packet. The Gaussian packet can be identified by using a dynamical potential barrier, with a resolution reaching the Heisenberg minimal uncertainty $\hbar/2$.
\end{abstract}
\maketitle


On-demand single-electron sources have been developed, opening a road towards a fermion version of quantum optics and related quantum processing~\cite{Pekola2013}. They are formed in a two-dimensional electron gas (2DEG), and can be classified according to emission energy. A mesoscopic capacitor~\cite{Feve2007} and a Leviton pump~\cite{Levitov2006, Glattli2013} emit electrons at Fermi energy, constituting one class. Significant steps, including the generation of a prescribed state~\cite{Levitov2006,Glattli2013}, Hanbury Brown-Twiss effects~\cite{Feve2012},  and single-electron quantum state tomography~\cite{Glattli2014,Grenier}, have been experimentally and theoretically done for this class~\cite{Buttiker2008,Buttiker2011,Feve2013,Buttiker2008-2}.

A quantum-dot (QD) pump~\cite{Blumenthal2007,Giblin2010,Slava2010,Masaya2011,Giblin2012,Fletcher2012,Masaya2013, Fricke2013,Kim2014,Slava2014,Haug2014,Waldie2015,Masaya2015} belongs to another class. It emits hot electrons of $\sim 100$ meV above the Fermi energy and has been studied for metrology~\cite{Giblin2012, Kim2014}. 
Its applications to fermion optics are complementary to those of the former class, and plausible as scattering of the hot electrons by phonons and other electrons rarely occurs~\cite{Taubert2011,Masaya2013,Emary2015}. 
Controllability of its emission energy is an important merit absent in the first class. This direction to fermion optics has been considered only recently~\cite{Haug2014}.

For various purposes of the direction, it is crucial to realize a QD pump emitting a prescribed electron wave packet of a useful form.
How to pump such states is a nontrivial question that requires an understanding of the latter half (emission process) of one pump cycle.
This process has been little considered theoretically, while the first half (capturing) was analyzed~\cite{Slava2010} for pump accuracy. 
State evolution in the emission involves complications from nonadiabatic operations 
and tunneling through a QD barrier;
how the shape of a wave packet changes by tunneling is a basic question that has not been addressed. As we show below, the change is governed by competition between Landauer-Buttiker traversal time~\cite{Buttiker1982, Landauer1994} $\tau_\textrm{tun}$, a characteristic scale of tunneling, and passage time~\cite{Tamm1945} $\tau_p$ for a packet to evolve into an orthogonal state.
In addition, a detector for measuring an emitted packet is crucial for fermion optics. It has not been analyzed, although it was addressed~\cite{Masaya2013,Waldie2015}.

In this Letter, we show, analytically and numerically with realistic parameters, that  a single-electron Gaussian packet can be emitted, without fine-tuning, from a QD pump under a strong magnetic field. 
When an electron is adiabatically captured in the QD, it then evolves into a coherent state, 
insensitive to details of the emission process, provided that the magnetic field is sufficiently strong. 
When the coherent state is emitted through a QD barrier, it becomes a single Gaussian packet 
in the regime of $\tau_\mr{tun} \ll \tau_p$ and a series of log-logistic packets~\cite{Kantam} for $\tau_\mr{tun} \gg \tau_p$.
The emitted state 
can be experimentally identified by using a dynamical potential barrier, with a resolution reaching the minimal uncertainty $\hbar/2$. 

\begin{figure}[b]	
\includegraphics[width=0.83\columnwidth]{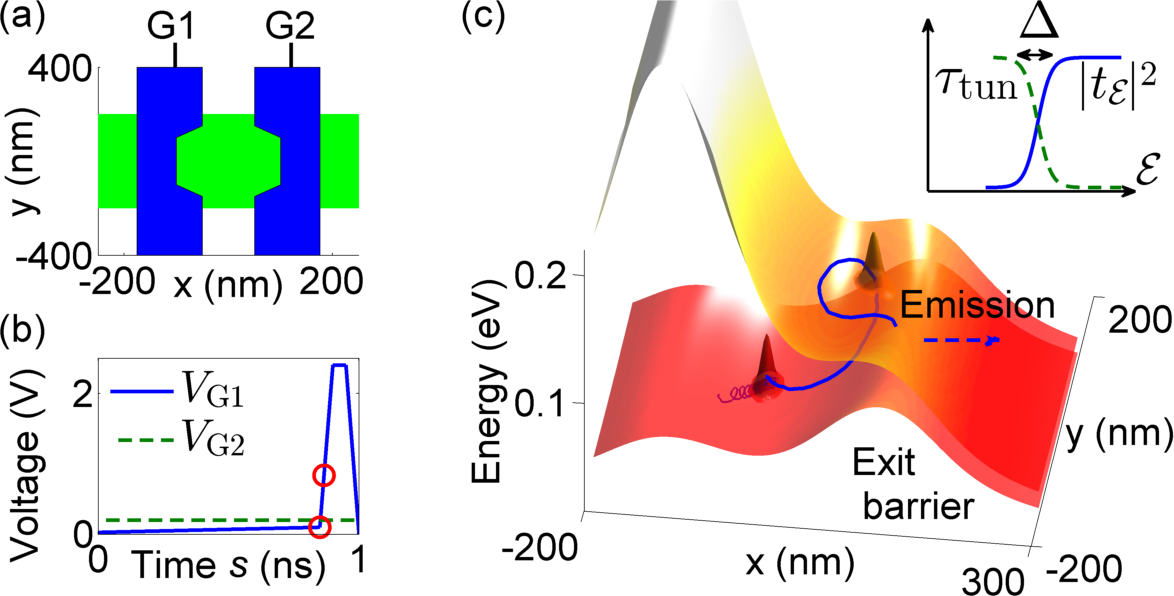}
\caption{
QD pump with realistic parameters~\cite{Masaya2013}.
(a) Gates $G1$ and $G2$ (blue) with voltages $V_{G1}$ and $V_{G2}$ form entrance and exit barriers in a GaAs 2DEG (green) located at 70~nm below. A magnetic field of 14 T is applied.
(b) $V_{G1}$ changes adiabatically and then abruptly in the emission process. $V_{G2}$ is fixed. 
(c) Numerical result at the times $s$ denoted by circles in (b). The ground state at $s=0$ evolves to a coherent state (cone) moving along the $\mb{E} \times \mb{B}$ drift (curve) as the barriers rise. Inset: Schematic plot of $\tau_\textrm{tun}(\mathcal{E})$ and $|t_\mathcal{E}|^2$.}
\label{fig1}
\end{figure}

We emphasize our general findings.
First, for electron dynamics in a time-dependent confinement potential in 2D, a strong perpendicular magnetic field $\mb{B} = B \mb{\hat{z}}$ is as useful as adiabatic and sudden regimes:
A Gaussian packet of width $\simeq l_B$ evolves into a coherent state moving along the classical $\mb{E} \times \mb{B}$ drift, as long as the confinement has a size much larger than magnetic length $l_B = \sqrt{\hbar/(|e| B)}$ and changes slowly in time scale $\omega^{-1}_c$. Here $\omega_c = |e|B / m_e^*$ is the cyclotron frequency of electron charge $e$ and mass $m_e^*$. Second, wave-packet tunneling dynamics is governed by $\tau_\mr{tun}$ and $\tau_p$. The character of a packet (except its velocity) is preserved in tunneling when $\tau_\mr{tun} \ll \tau_p$, while it is lost and the packet splits into partial waves of different energy when $\tau_\mr{tun} \gg \tau_p$.

{\it Coherent state in a QD pump.---}A QD pump is formed in a 2DEG  by gate voltages $V_{G1}$ and $V_{G2}$ and driven by modulating $V_{G1}$; see Fig.~\ref{fig1}. In the capturing process of each cycle, electrons are confined in the QD. In the emission process, they are emitted through the exit barrier, as the entrance barrier becomes sufficiently large.
We focus on the regime that only one electron is captured in the QD and occupies the ground state in the capturing. This regime occurs when $V_{G1}$ adiabatically changes at a low temperature~\cite{NOTE1}.
Recent experiments~\cite{Masaya2011} demonstrated a single-electron capture, with suppressing 
excitations in the capturing process.


Time evolution of the captured electron in the emission process is not trivial, since the QD confinement potential $U_\textrm{QD}$ nonadiabatically changes in time $s$. 
Perhaps surprisingly, a strong magnetic field 
makes the evolution simple and insensitive to process details. 
At a time $s = 0$ of the capturing process, the electron is in the ground state of the QD. 
When 
$\omega_c \gg$ 
the QD confinement frequency $\omega_0$, the ground state is Gaussian. 
To see its evolution, we consider a Hamiltonian  
$H_\textrm{QD} = (\mb{p}- e\mb{r}\times B \hat{\mb{z}}/2)^2 / (2m_e^*)  +  U_\textrm{QD}(\mb{r}, s)$, where $\mb{p}$ and $\mb{r}$ are electron momentum and position, respectively. 
We find~\cite{Supple} that its time evolution $\psi (\mb{r},s)$ is well described by 
\begin{equation}
 \psi \simeq \psi_c(\mb{r}, s) = \frac{1}{\sqrt{\pi l_B^2}} \exp \left[ i \mb{r} \cdot \mb{p}_c (s)   - \frac{[\mb{r}-\mb{r}_c(s)]^2}{2 l_B^2} \right], \label{coh_state} \end{equation}
provided that $U_\textrm{QD}$ changes slowly in length scale $l_B$ and time scale $\omega_c^{-1}$. We call $\psi_c$ a coherent state, as
$\mb{r}_c (s)$ and $\mb{p}_c (s)$ follow the classical $\mb{E} \times \mb{B}$ drift governed by $\dot{\mb{r}} = \partial H_\textrm{QD} / \partial \mb{p}$ and $\dot{\mb{p}} = - \partial H_\textrm{QD}/ \partial \mb{r}$. 

$\psi (\mb{r},s)$ becomes identical to the coherent state, when $\omega_c \gg \omega_0$ and the anisotropy of $U_\textrm{QD}(\mb{r}, s)$ is not too large. 
For example, we consider $U_\textrm{QD} = [\omega_{0,x}^2(s) x^2+\omega_{0,y}^2(s) y^2]/(2 m_e^*) +F(s) x$, where the force $F(s)$ describes a shift of the harmonic confinement center in time. 
In this case,  we find~\cite{Supple} $|\ovl{\psi_c(\mb{r}, s)}{\psi(\mb{r}, s)}|^2 \approx 1 -   \omega_0^4/ \omega_c^4 -(\omega_{0,x}-\omega_{0,y})^2/(8\omega_0^2)$, where $\omega_0(s) = [\omega_{0,x}(s)+\omega_{0,y}(s)]/2$.
For the realistic case $U_\textrm{QD}(\mb{r}, s)$ of Fig.~\ref{fig1}, we numerically confirm $|\ovl{\psi_c(\mb{r}, s)}{\psi(\mb{r}, s)}|^2 \approx 1$, solving the Laplace equation~\cite{Sukhorukov1995} for $U_\textrm{QD}(\mb{r}, s)$ with the boundary condition by $V_{G1,G2}$; 
we estimate $\hbar \omega_0 (s) \simeq 3$~meV, $\hbar \omega_c = 24$~meV at $B = 14$~T, and $|\omega_{0,x}(s)-\omega_{0,y}(s)| / \omega_0(s) \sim 0.3$, leading to $\omega_0^4/\omega_c^4 \lesssim 10^{-4}$, $(\omega_{0,x}-\omega_{0,y})^2/(8\omega_0^2) \lesssim 10^{-2}$, and $(2\pi / \omega_c) \partial\omega_{0}/(\omega_0 \partial s) \lesssim 10^{-3}$; notice that the anisotropy of $|\omega_{0,x}-\omega_{0,y}| / \omega_0 \sim 0.3$ negligibly affects $|\ovl{\psi_c(\mb{r}, s)}{\psi(\mb{r}, s)}|^2$.
Remarkably, $\psi (\mb{r},s)$ is characterized only by $l_B$ and the $\mb{E} \times \mb{B}$ drift.

{\it Time scales.---}The emission dynamics of the coherent state through the exit barrier
is governed by the time scales of the barrier and the state.
The exit barrier at a given time is modeled as a saddle potential of 
$U_\textrm{saddle} -  m_e^* (\omega_{b,x}^2 x^2 - \omega_{b,y}^2 y^2)/2$ with frequencies $\omega_{b,x/y}$ and constant $U_\textrm{saddle}$. A plane wave of energy $\mathcal{E}$, moving along an equipotential line in the QD, has transmission amplitude $t_\mathcal{E}$ through the barrier~\cite{Buttiker1990}:
$|t_\mathcal{E}|^2 = 1 / [1 + e^{- 2 \pi ( \mathcal{E} - U_\textrm{saddle} - \hbar \omega_c / 2 ) / \Delta)}]$.
$\Delta \equiv \hbar \omega_{b,x} \omega_{b,y} / \omega_c$ is the energy window where $|t_\mathcal{E}|^2$ varies with $\mathcal{E}$ and has the same order with QD level spacing $\hbar \omega_0^2 / \omega_c$ as $\omega_{b,x/y} \simeq \omega_0$ typically.
The traversal time $\tau_\mr{tun} \equiv \hbar \partial \ln t_\mathcal{E}/\partial \mathcal{E}$, during which the plane wave passes through the barrier, is expressed~\cite{Buttiker1982, Landauer1994} as (see Fig.~\ref{fig1})
\begin{equation} \label{Ttun}
\tau_\mr{tun}(\mathcal{E}) = \frac{\pi \hbar}{\Delta} \frac{1}{ 1+ e^{2\pi (\mathcal{E} - U_\textrm{saddle} - \hbar \omega_c/2)/\Delta} }.
\end{equation}
The barrier is characterized by
$\Delta$, $\tau_\mr{tun}$, and the speed $v_U$ of its height change by $V_\textrm{G1}(s)$.


The coherent state with drift speed $v$ has passage time~\cite{Tamm1945} $\tau_p = \sqrt{2} l_B / v$ for it to evolve into an orthogonal state. 
It has travel time $\tau_d \simeq 2 \pi \omega_c / \omega_0^2$ for circling once along the QD circumference. After emitted, the state has an energy distribution with peaks separated by $v_U \tau_d$ within a window of $\Delta$, since its transmission probability through the barrier increases whenever it arrives at the barrier.

In experiments~\cite{Giblin2012}, $v_U \tau_d \gg \Delta$. We estimate $v_U \sim 1$~meV/ps, as $V_{G1}$ changes with $\sim 10$ mV/ps in the emission~\cite{NOTE2}. And for the setup in Fig.~\ref{fig1}, $\Delta \sim 0.4$ meV, $\tau_d \sim 10$ ps, and hence $v_U \tau_d / \Delta \sim 25$.

{\it Emission of a coherent state.---}  As $V_{G1}$ changes, the coherent state propagates along the $\mb{E} \times \mb{B}$ drift around the QD circumference. Whenever it approaches with a sufficiently large energy to the barrier, it can be emitted out of the QD and then move along the 2DEG edge.

\begin{figure}[tb]
\includegraphics[width=0.85\columnwidth]{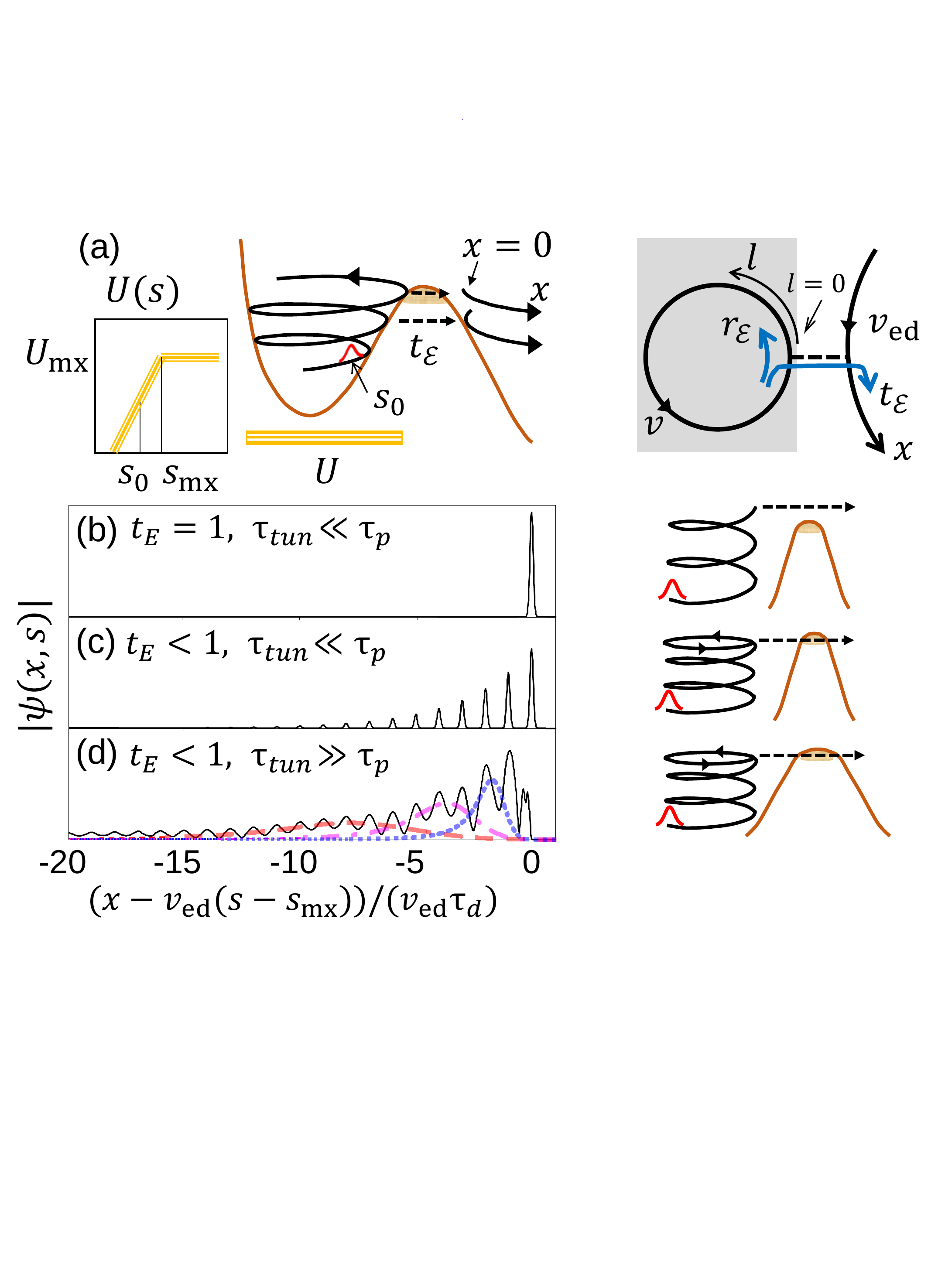}
\caption{(a) Trajectory (solid arrows) and emission (dashed arrows) of a coherent state. Right: The QD is modeled by a loop.
(b)-(d) An emitted state $\psi(x,s)$ obtained from Eqs.~\eqref{ga} and \eqref{exp}, and the corresponding dynamics. It is (b) a single Gaussian packet, (c) a series of Gaussian packets, or (d) a series of log-logistic packets.
Dashed, dotted, and dash-dot curves in (d) show three dominant $\textrm{sech}$ functions in Eq.~\eqref{exp}. 
}
\label{fig2}
\end{figure}   

We derive the emitted state for $v_U \tau_d \gtrsim \Delta$, applying the following model (see Fig.~\ref{fig2}) to the coherent state [Eq.~\eqref{coh_state}] after a time $s_0$ that the coherent state starts to be emitted through the barrier:
The coherent state circles along a loop with constant velocity $v$, gaining spatially uniform potential energy $U(s)$ by $V_{G1}$. $U(s)$ increases with $s$ and then stays at its maximum $U_\textrm{mx}$ at $s > s_\textrm{mx}$, following $V_{G1}(s)$ in Fig.~\ref{fig1}.
The exit barrier, located at $l=0$, is characterized by amplitude $r_\mathcal{E}$ ($t_\mathcal{E}$) with which the plane wave of energy $\mathcal{E}$ is reflected by (emitted through) the barrier.
At time $s_0$, the coherent state is located at $l=0$ for brevity and has so low energy that $t_\mathcal{E} = 0$ within its energy window.
The model is valid for $v_U \tau_d \gtrsim \Delta$, since $|t_\mathcal{E}|^2$ varies from 0 to its maximum in the time during which the state circles only once around the QD circumference.   
We apply a time-dependent scattering theory~\cite{Supple}: (i) We attach the dynamical phase by $U(s)$ to scattering amplitudes at $l=0$ via gauge transformation as in a Floquet theory~\cite{floquet}, (ii) derive Fabry-Perot scattering states using the plane waves, and (iii) obtain the emitted state, computing the overlap between the coherent state and the scattering state at $s_0$.


The resulting emitted wave function is written as
\begin{eqnarray} 
\psi(& &s_\textrm{rd},y) = Y\hspace{-.1cm} \sum_{n,m=1}^{\infty} \hspace{-.1cm}  e^{-(\mathcal{E}_n -\epsilon_0)^2 \tau_p^2/4} \,
t_{E^{nm}_{m}} 
\prod_{m'=0}^{m-1} r_{E^{nm}_{m'}}  \nonumber\\
& & \times  e^{ i m \mathcal{E}_n \tau_d/\hbar + im \phi_\textrm{AB} -i \int_{s_0}^{s_\textrm{rd}} { [\mathcal{E}_n + U (s')] ds' / \hbar }}  \zeta_m (s_\mathrm{rd}).   \label{ew}
\end{eqnarray}
$s_\textrm{rd} \equiv s - \frac{ x}{v_\textrm{ed}} > 0$ is the retarded time by electron velocity $v_\textrm{ed}$ along the 2DEG edge with longitudinal coordinate $x$ (transverse $y$), $Y \propto e^{- y^2 / (2l_B^2)}$, $\zeta_m = 1$ (0) for $m \le (s_\textrm{rd} - s_0)/\tau_d < m+1$ 
(otherwise), and $\epsilon_0$ is the energy of the coherent state at $s_0$.
The $(n,m)$ term of $\psi$ describes an electron that
occupies the $n$th QD resonance state of kinetic energy $\mathcal{E}_n$
at $s_0$ and is emitted after $m$ circulations with amplitude $t_{E^{nm}_{m}} \prod_{m'=0}^{m-1} r_{E^{nm}_{m'}}$; each circulation leads to dynamical phase $\mathcal{E}_n \tau_d / \hbar$ and Aharanov-Bohm phase $\phi_\textrm{AB}$; the resonance obeys $\mathcal{E}_n \tau_d / \hbar + \phi_\textrm{AB} = 2 \pi n$.
This electron has 
energy $E^{nm}_{m'} \equiv {\mathcal{E}_n} +U(s_\textrm{rd} -(m-m') \tau_d)$ 
at the $m'$th reflection ($m' < m$) before its emission; energies are hereafter measured relative to $U(s_0)$. 
$e^{-({\mathcal{E}_n} -\epsilon_0)^2 \tau_p^2/4}$
is the overlap weight between the $n$th resonance and the coherent state.
When $U$ is time independent, 
Eq.~\eqref{ew} describes a usual Fabry-Perot state.


We analyze $\psi$ in the case of $\tau_\mr{tun} \ll \tau_p$.
In this case, $t_{E^{nm}_{m}}$ and $r_{E^{nm}_{m'}}$ are $n$ independent in energy window $\hbar/\tau_p$, 
$t_{E^{nm}_{m}} \simeq t_{\epsilon_0 + U(s_\textrm{rd})}$ and $r_{E^{nm}_{m'}} \simeq r_{\epsilon_0 +U(s_\textrm{rd} -(m-m')\tau_d)}$.
Hence, the sum over $n$ in Eq.~\eqref{ew} is done:
\begin{eqnarray} 
& &  \psi(s_\textrm{rd} > s_0,  y)  \propto Y t_{\epsilon_0 + U(s_\textrm{rd})}  \sum_{m=1}^{\infty}   \prod_{m'=0}^{m-1} r_{\epsilon_0 +U(s_\textrm{rd} -(m-m')\tau_d)}  \nonumber \\
& & \,\,\, \times e^{ -\frac{i}{\hbar} \epsilon_0(s_\textrm{rd} -s_0 - m \tau_d) }
e^{- \frac{i}{\hbar} \int_{s_0}^{s_\textrm{rd}} U(s')ds'} e^{- \left(\frac{s_\textrm{rd} -s_0 -m\tau_d}{ \tau_p}\right)^2}. \label{ga}
\end{eqnarray}
$\psi$ is a series of Gaussian packets with positions separated by $v_\textrm{ed} \tau_d$ and amplitudes decreasing sequentially. 
Remarkably, when $\tau_\mr{tun} \ll \tau_p$, the emission instantly occurs and
the character of the coherent state is preserved.

A single Gaussian packet of $e^{- (s_\textrm{rd} - s_0 -\tau_d)^2/\tau_p^2}$ can be emitted in a range of $V_{G2}$, when $\tau_\mr{tun} \ll \tau_p$. Under the condition of $v_U \tau_d \gg \Delta$ (achieved in experiments~\cite{Giblin2012}),
one controls $V_{G2}$ to make the emission occur only at $s > s_\textrm{mx}$ with $|t_{\epsilon_0 + U_\textrm{mx}}| = 1$.
Then, a single Gaussian packet is emitted as in Fig.~\ref{fig2}(b).
Note that if $|t_{\epsilon_0 + U_\textrm{mx}}| < 1$, a series of Gaussian packets are generated 
as in Fig.~\ref{fig2}(c).  
 
\begin{figure}[b] 
\includegraphics[width=0.68\columnwidth]{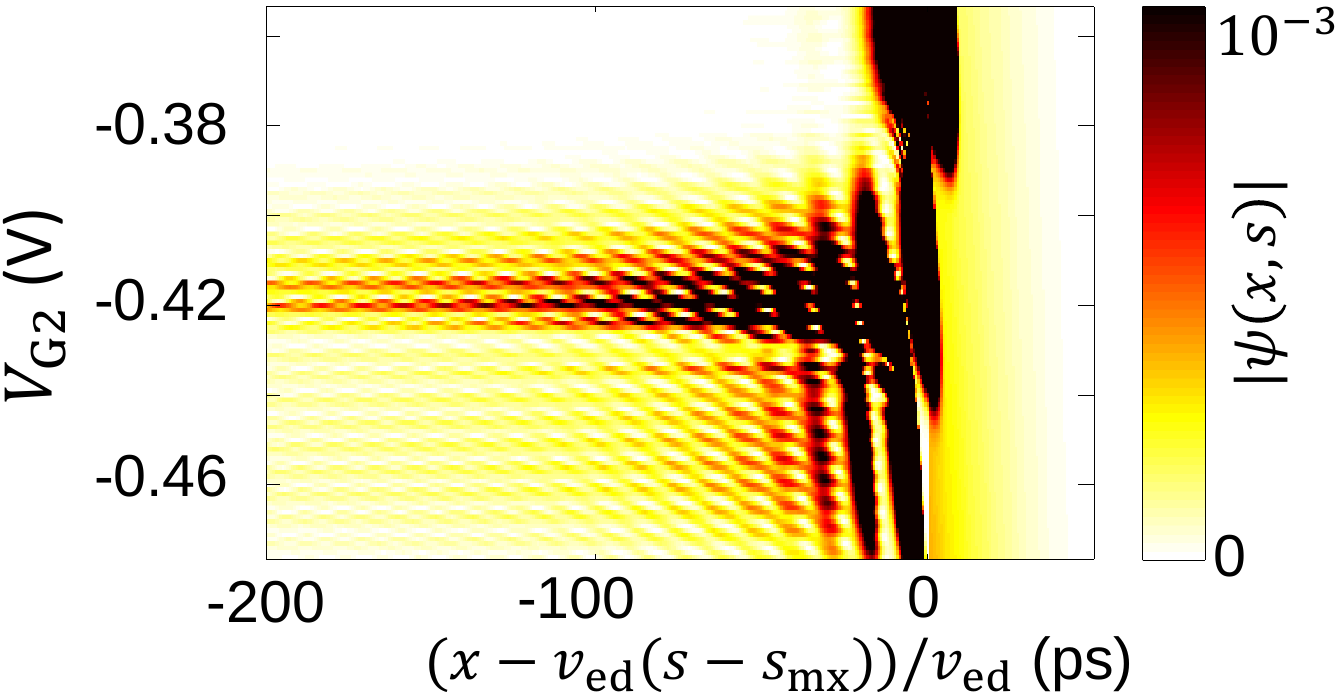}
\caption{ Numerical results of an emitted state $|\psi(x,s)|$ versus $V_{G2}$. It is a Gaussian packet at $V_{G2}$ $\gtrsim -0.38$~V and a series of log-logistic functions at smaller $V_{G2}$.
}
\label{fig3}
\end{figure}

In the opposite limit $\tau_\mr{tun} \gg \tau_p$, the coherent state loses its character in the emission, splitting into resonance components.
This limit occurs when $|t_{\epsilon_0 + U_\textrm{mx}}|$ is small. 
Hence, the emission occurs mostly during the time interval of $U(s) = U_\textrm{mx}$  
as $|t_\mathcal{E}|$ is smaller before the time. 
The time interval 
is 100 ps typically~\cite{Giblin2012}, larger than $\tau_d \sim 10$ ps. 
In this time, QD resonance levels have broadening $|t_{ { \mathcal{E}_n} + U_\textrm{mx}}|^2 \hbar / \tau_d$ much narrower than 
level spacing $\hbar / \tau_d$, 
and their transmission amplitude $|t_{ { \mathcal{E}_n} + U_\textrm{mx}}|$ is a rapidly increasing function of $n$, since $\tau_\mr{tun} \gg \tau_p$.
Hence, the resonance levels constituting the coherent state are emitted one by one; a larger-$n$ level is emitted earlier around its lifetime $s_\textrm{rd} \simeq \tau_d / |t_{{  \mathcal{E}_n} + U_\textrm{mx}}|^2$.
Indeed, Eq.~\eqref{ew} is reduced~\cite{Supple} into a sequence of log-logistic wave functions:
\begin{eqnarray}
\psi (& & s_\textrm{rd}, y) \propto \, Y \sqrt{\frac{2\tau_d}{|t_{\epsilon_0 + U_\textrm{mx}}|^2 s_\textrm{rd}}} e^{-  (\frac{\tau_p}{\tau_\mr{tun}} \ln \frac{|t_{\epsilon_0 + U_\textrm{mx}}|^2 s_\textrm{rd}}{\tau_d})^2}  \nonumber \\
& & \times  \sum_{n=1}^{\infty} e^{i ({  \mathcal{E}_n} + U_\textrm{mx}) \frac{s_\textrm{rd}}{\hbar}} \mr{sech} (\pi \ln  \frac{|t_{{  \mathcal{E}_n} + U_\textrm{mx}}|^2 s_\textrm{rd}}{2\tau_d} ), \label{exp}
\end{eqnarray}
where $s_\textrm{rd}$ $(>0)$ is measured relative to $s_\textrm{mx}$.
In $s_\textrm{rd} \in$ [$\tau_d/|t_{{  \mathcal{E}_n}  + U_\textrm{mx}}|^2$, $\tau_d/|t_{{  \mathcal{E}_{n-1}} + U_\textrm{mx}}|^2$],
two resonances $n-1$ and $n$ dominantly contribute to $\psi$, showing interference  $|\psi (s_\textrm{rd})| \propto | ( |t_{{  \mathcal{E}_n} + U_\textrm{mx}}|^2 s_\textrm{rd} / \tau_d )^{-\pi} + ( |t_{{  \mathcal{E}_{n-1}} + U_\textrm{mx}}|^2 s_\textrm{rd} / \tau_d )^{\pi} e^{2\pi i s_\textrm{rd} / \tau_d}|$ with oscillation period $\tau_d$; see the sech function and Fig.~\ref{fig2}(d).

In Fig.~\ref{fig3}, we numerically compute $|\psi(s_\textrm{rd})|$ for the setup (including the 2D outside the QD) in Fig.~\ref{fig1}. 
It shows a single Gaussian packet at $V_\textrm{G2} > -0.38$ V and a series of log-logistic packets at $V_{G2} < -0.41$ V in agreement with Eqs.~\eqref{ga} and \eqref{exp}; we estimate $\tau_p \simeq 1.4$ ps, $\tau_\mr{tun}(\epsilon_0 + U_\textrm{mx}) \simeq 11$ ps at $V_{G2}$ = -0.41 V, and $\tau_\mr{tun}(\epsilon_0 + U_\textrm{mx}) \simeq 0.04$ ps at $V_{G2}$ = -0.4 V.
This shows experimental possibility of generating a prescribed state by varying $V_{G2}$. 
Note that $|\psi(s_\textrm{rd})|$ is weaker and shifted by $\tau_d$ at $V_{G2}$ below -0.43 V than above, since the coherent state is emitted after one more circulation due to the higher exit barrier.

The interference pattern at $V_{G2} < -0.41$ V in Fig.~\ref{fig3}, described in Eq.~\eqref{exp}, can be experimentally measured by using a dynamical potential barrier (discussed later) or from electron current $I_p$ by the QD pump.
In the latter, the emitted state is linked to
$I_p = e f \int_{0}^{1/f} |\psi(x=x_D, s)|^2 ds$,
where $x_D$ is the detector position and $f$ is the pumping frequency. $\partial (I_p/ef)/\partial f$ shows the pattern.

\begin{figure}[t] 
\includegraphics[width=0.95\columnwidth]{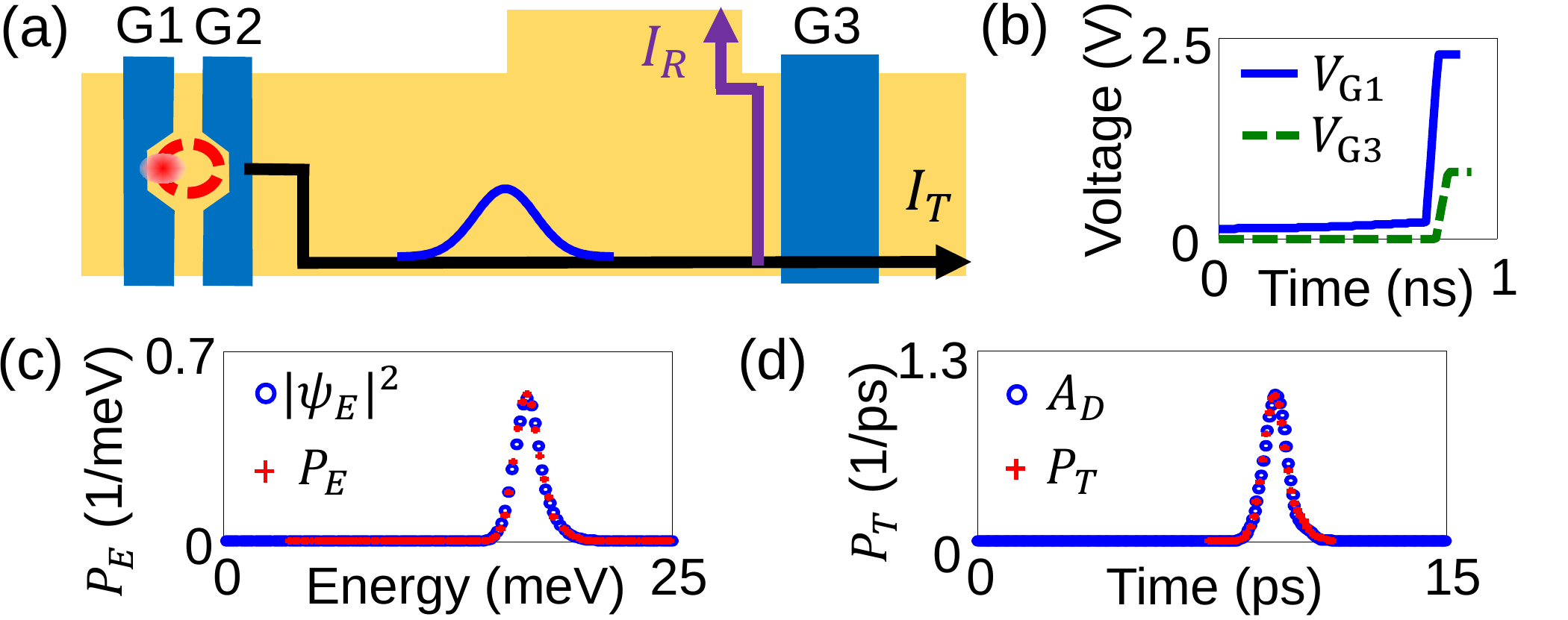}
\caption{
(a) A Gaussian wave packet is emitted from the pump in Fig.~\ref{fig1}. Its energy ($P_E$) and arrival-time ($P_T$) distributions are detected by using gate $G3$ of 200 nm width and voltage $V_{G3}$.
(b) Time dependence of $V_{G3}$ for measuring $P_T$. $V_{G3}$ changes 1 V in 50 ps. 
(c),(d) Numerical results of $P_E(E)$ and $P_T(s)$, compared with $|\psi(\mathcal{E})|^2$ and $A_D(s)$.
}
\label{fig4}
\end{figure}

{\it Detection of a Gaussian state.---}An emitted state can be identified by measuring current $I_T$ through a potential barrier induced by gate $G3$ (Fig.~\ref{fig4}), as shown experimentally~\cite{Masaya2013, Waldie2015}. As below, 
it is useful for studying the emission dynamics of $\tau_\textrm{tun}$ and $\tau_p$ with a resolution reaching the Heisenberg minimal uncertainty $\hbar/2$.

The energy distribution $|\psi(x_D, \mathcal{E})|^2$ of the emitted state [the Fourier transformation of  $\psi(x_D, s)$]
is obtained from 
$P_E (U_3) \equiv (e f)^{-1} \partial I_T / \partial U_3$,
by measuring $I_T$ with time-independent barrier height $U_3$.
$I_T$ is written as  $I_T = ef \int d \mathcal{E} |\psi(x_D, \mathcal{E})|^2 T_3(\mathcal{E};U_3)$. $T_3(\mathcal{E};U_3)$ is the transmission probability at energy $\mathcal{E}$ through the $G3$ barrier of height $U_3$.
One has better resolution as $\partial T_3(\mathcal{E};U_3) / \partial U_3 \to \delta(\mathcal{E} - U_3)$. 
We note that $\int d U_3 P_E (U_3) = 1$, and a method for converting the value of $V_{G3}$ to $U_3$ is known~\cite{Taubert2011}.
For the square barrier with height $U_3$ and width $L$, $T_3(\mathcal{E}; U_3) \approx \exp(-0.083\sqrt{\frac{U_3 - \mathcal{E}}{\mathrm{meV}}}\frac{L}{\mathrm{nm}} )$ and the width of $\partial T_3(\mathcal{E};U_3)/\partial \mathcal{E}$ is $150 \left(\frac{\mathrm{nm}}{L} \right)^2 \mathrm{meV}$; it is obtained by the WKB method.
This means an energy resolution $\sim 10^{-3}$ meV for $L= 200$ nm,
which is much smaller than the energy width $\hbar / \tau_p \sim 0.66$ meV of the Gaussian packet.

The arrival time distribution of the emitted state is defined~\cite{Muga2000}
as $A_D(s) \equiv \frac{\partial}{\partial s} \int_{x_D}^{\infty} dx |\psi(x_D, s)|^2$; it is similar to the waiting time distribution~\cite{Albert2011}.
To obtain $A_D(t)$, one rapidly raises $U_3$ from 0 to a large value ($\gg$ the energy of the emitted state) at time $s_3$ and measures  
$P_T(s_3) \equiv (ef)^{-1} \partial I_T / \partial s_3$.
It approaches to $A_D(s)$ under the condition  $\hbar / (v_3 \tau_p^2) \ll 1$ where the time uncertainty $\tau_p$ of the Gaussian state is much larger than the rise time $\hbar / (v_3 \tau_p)$ of $U_3$ (with speed $v_3$ [energy/time]) over the energy uncertainty $\hbar / \tau_p$. 
When $V_{G3}$ changes by 1 V in 50 ps,  $v_3 \sim 10$ meV/ps \cite{NOTE2} and the condition is satisfied as $\hbar / (v_3 \tau_p^2) \sim 0.03$; $\tau_p \sim 1.4$ ps at $B=14$ T.


In Fig.~\ref{fig4}, a whole process from the generation to the detection of a Gaussian packet is numerically simulated.
The product of the resulting peak widths of $P_T$ and $P_E$ reaches about 1.1 times $\hbar / 2$, which well matches (within an error $\sim 1/100$) with the energy-time uncertainty directly obtained from the numerical result of $\psi(x,s)$. The deviation from $\hbar/2$ is due to the finite change speeds of $V_{G1}$ and $V_{G3}$; the former can generate additional small packets in Eq.~\eqref{ga}.
In experiments, there can appear additional sources of deviation:
The coherent state can be emitted into multiple Landau levels of the 2DEG (resulting in a non-Gaussian packet) rather than only into the lowest one, which is negligible at $B \gtrsim 10$ T where the spatial Landau-level separation $\gg l_B$; scattering of an emitted packet by phonons and other 2DEG electrons is also negligible, as the mean free path of the packet is longer than 5 $\mu$m and tunable to be longer~\cite{Taubert2011, Masaya2013,Emary2015}.



{\it Conclusion.---}We found general dynamical properties (coherent-state motion and tunneling dynamics) of an electron wave packet in a 2D dynamic quantum dot under a strong magnetic field and demonstrated that a Gaussian packet can be generated and detected.  We emphasize that almost the same Gaussian packet can be generated from nonidentical QD pumps.  
The generation will be useful for the experimental study of packet tunneling dynamics ($\tau_\textrm{tun}$ versus $\tau_p$), testing the efficiency of a wave-function detector (whether reaching the $\hbar/2$ resolution limit), and investigating two-electron Hanbury Brown-Twiss correlations (where two identical incident states are necessary) and two-electron Coulomb collision.
The minimal uncertainty of the packet can be experimentally confirmed by using a dynamical barrier.
It will be valuable to further develop $P_E$ and $P_T$ to the level of the single-electron quantum state tomography~\cite{Glattli2014,Grenier}.


We thank Clive Emary and Vyacheslavs Kashcheyevs for valuable discussions. 
H.S.S. is supported by Korea NRF (Grant No. 2015R1A2A1A15051869 and No. 2016R1A5A1008184).
M.K. is supported by the United Kingdom Department for Business, Innovation and Skills.
This work is supported by the Brain Korea 21 PLUS Project of Korea Government.

\pagebreak
\vspace{.3cm}
\widetext
\begin{center}
\textbf{\large Supplemental Materials: Ultrafast Emission and Detection of a Single-Electron Gaussian Wave Packet: A Theoretical Study}
\end{center}
\setcounter{equation}{0}
\setcounter{figure}{0}
\setcounter{table}{0}
\setcounter{page}{1}
\makeatletter
\renewcommand{\theequation}{S\arabic{equation}}
\renewcommand{\thefigure}{S\arabic{figure}}
\renewcommand{\bibnumfmt}[1]{[S#1]}
\renewcommand{\citenumfont}[1]{S#1}

\section{Coherent state in a QD pump under a strong magnetic field}
We show that in a 2D QD formed by time-dependent confinement potential, the evolution of a Gaussian state with spatial width $\simeq l_B$ 
follows the classical equation of motion of a coherent state 
under the strong-magnetic-field conditions that the QD has a size much larger than the magnetic length $l_B$ and that the QD confinement potential $U_\textrm{QD}$ changes slowly in the length scale of $l_B$ and the time scale of $\omega_c^{-1}$.
We derive Eq.~(1) and the expression of the wave function overlap $|\ovl{\psi_c}{\psi}|^2$ mentioned in the main text.    
In order to elucidate the behavior and the validity of the result, we discuss an example, the time evolution of the ground state in a time-dependent anisotropic harmonic QD confinement potential.

We consider the ground state of the Hamiltonian $H_\textrm{QD}$ at initial time $s=0$, and study its time evolution under $H_\textrm{QD}(s) = (\mb{p}- e\mb{r}\times B \hat{\mb{z}}/2)^2 / (2m_e^*)  + U_\textrm{QD} (\mb{r}, s)$. Since at low energy the confinement potential $U_\textrm{QD} (s=0)$ is well approximated as an anisotropic harmonic potential, 
the ground state can be approximately expressed~\cite{Madhav-2} as a Gaussian form with certain harmonic frequencies $\omega_{0,x}$ and $\omega_{0,y}$,
\begin{equation} 
\psi (\mb{r}, s=0) \approx \exp[ -\{\frac{x^2}{ l_B^2}\frac{\omega_{0,x}}{\omega_{0,x} +\omega_{0, y}}+\frac{y^2}{ l_B^2}\frac{\omega_{0,y}}{\omega_{0,x} +\omega_{0, y}}\}(1+ \frac{(\omega_{0,x} +\omega_{0, y})^2}{2\omega_c^2}) +i \frac{xy}{l_B^2} \frac{\omega_{0,x}-\omega_{0,y}}{\omega_{0,x} +\omega_{0, y}} (1+\frac{(\omega_{0,x} +\omega_{0, y})^2}{{ 4} \omega_c^2})], \label{init_state} \end{equation}
up to the normalization constant. 
This approximation becomes more valid as $\omega_c$ becomes much larger than $\omega_{0,x}$ and $\omega_{0,y}$; when $\omega_c \gg \omega_{0,x}, \omega_{0,y}$,
$\psi (\mb{r}, s=0) \to \exp [ - \mb{r}^2 / (2 W) ]$ with $W \sim l_B$ and the overlap between the exact ground state and $\exp [ - \mb{r}^2 / (2 W) ]$ is approximately $1 - \bo (4 \omega_{0, x (y)}^4 / \omega_c^4)  $.

Next, we study the time evolution of the ground state $\psi (\mb{r}, s=0)$ in Eq.~\eqref{init_state} under $H_\textrm{QD}(s)$. 
The evolution is decomposed into two parts, one from the kinetic Hamiltonian and the other from $U_\textrm{QD}(\mb{r}, s)$, 
\begin{equation}\label{small_time_evol}
\psi(\mb{r'}, s + \delta s ) = \int d\mb{r} \, K_B(\mb{r'}, s + \delta s; \mb{r}, s) e^{-i {U_\textrm{QD}(\mb{r}, s)} \delta s +\bo[\delta s^2]}  \psi(\mb{r}, s)
\end{equation}
for infinitesimal $\delta s$. $K_B (\mb{r}', s; \mb{r},0) \equiv \langle \mb{r'}| e^{-i s (\mb{p}- e\mb{r}\times \mb{B}/2)^2 / (2m_e^*)} | \mb{r} \rangle$ is obtained~\cite{Papadopoulos1971-2} as 
\begin{equation}
K_B (\mb{r}', s; \mb{r}, 0) = \left(\frac{m_e^*\omega_c}{4\pi i \sin \frac{\omega_c s}{2}}\right)^{3/2} \exp\left[\frac{i}{2 l_B^2} \left( \cot \frac{\omega_c s}{2}(\mb{r}-\mb{r'})^2 +2 (\mb{r'} \times \mb{r})_z \right) \right]. \nonumber
\end{equation}
To compute the time evolution further, (i) the infinitesimal $\delta s$ is considered so that the commutator between the kinetic Hamiltonian and $U_\textrm{QD}(\mb{r}, s)$ is neglected in Eq.~\eqref{small_time_evol}, and (ii) at each instant $s$, $U_\textrm{QD}(\mb{r}, s)$ is approximately expressed as 
\begin{eqnarray}
U_\textrm{QD}(\mb{r}, s) \simeq U_\textrm{QD}(\mb{r}_0 , s) +\sum_{i=x,y} \frac{\partial}{\partial x_i} U_\textrm{QD}(\mb{r}, s)|_{\mb{r}_0} (\mb{r}-\mb{r}_0)_i +\sum_{i,j=x,y}\frac{\partial^2}{2\partial x_i \partial x_j} U_\textrm{QD}(\mb{r}, s)|_{\mb{r}_0} (\mb{r}-\mb{r}_0)_i (\mb{r}- \mb{r}_0)_j, \label{TaylorExp} \end{eqnarray}
where
$U_\textrm{QD}(\mb{r}_0 , s)$ is the potential at the mean position $\mb{r}_0$ of $\psi(\mb{r}, s)$ at time $s$. This Talyor expansion up to the second order is sufficient when $U_\textrm{QD}(\mb{r}, s) - U_\textrm{QD}(\mb{r}_0, s)$ is much smaller than $\hbar \omega_c$ for $|\mb{r}-\mb{r}_0| < l_B$;
when the total potential $U_\textrm{QD}(\mb{r}, s) + m_e^* \omega_c^2 \mb{r}^2 / 8$ from $U_\textrm{QD}$ and the magnetic confinement is Taylor expanded around $\mb{r}_0$, 
the expansion up to the second order dominates over the higher-order terms. Note that the drift motion of the state evolution is described by the first two terms of Eq.~\eqref{TaylorExp}, while the anisotropic shape of the Gaussian form is determined by the last term of Eq.~\eqref{TaylorExp}.
To compute the time evolution of the state, we apply Eq.~\eqref{TaylorExp} to Eq.~\eqref{small_time_evol} and perform Gaussian integrals, assuming that $U_\textrm{QD}$ changes slowly in time $< \omega_c^{-1}$, namely $\partial U_\textrm{QD} / \partial s \ll \omega_c U_\textrm{QD}$.   
The result shows that the time evolved state remains in a Gaussian form.
\begin{equation}\label{finite_time_evol}
\psi(\mb{r}, s) = N \exp\left[ (\mb{r}-\mb{r}_c(s))^{\intercal} { \mb{R}} (\mb{r}_c(s))^\dagger { \begin{bmatrix} A(s) & C(s)/2 \\ C(s)/2 & B(s) \end{bmatrix} } { \mb{R}} (\mb{r}_c(s)) (\mb{r}-\mb{r}_c(s)) /l_B^2 \right] 
 e^{i \mb{p}_c(s) \cdot \mb{r} }
\end{equation}
where $N \equiv \sqrt[4]{4 \mr{Re}[A(s)] \mr{Re}[B(s)] -\mr{Re}[C(s)]^2}/(\sqrt{\pi}l_B)$. 
The matrix $\mb{R}(\mb{r}_c(s)) \equiv \begin{bmatrix} \cos \phi(\mb{r}_c(s)) & \sin \phi(\mb{r}_c(s)) \\ -\sin \phi(\mb{r}_c(s)) & \cos \phi(\mb{r}_c(s)) \end{bmatrix}$ rotates the coordinate by  angle $\phi(\mb{r}_c(s)) \equiv \frac{1}{2} \tan^{-1}[ \partial_{xy}^2 U_\textrm{QD}(\mb{r}_c(s))/\{\partial_{yy}^2 U_\textrm{QD}(\mb{r}_c(s))- \partial_{xx}^2 U_\textrm{QD}(\mb{r}_c(s))\}]$ so that the anisotropic directions (the major and minor axes) of the Gaussian form align the  rotated axes $\mb{R}(\mb{r}_c(s)) \hat{\mb{x}}$ and $ \mb{R}(\mb{r}_c(s))  \hat{\mb{y}}$.
{ 
Remarkably, We notice that the mean position $\mb{r}_c(s)$ and mean momentum $\mb{p}_c(s)$ of the state at time $s$ are determined by the classical equation of motion (this is why we call $\psi$ a coherent state),
\begin{equation} \label{eom_rp}
\begin{aligned}{}
\frac{d \mb{r}_c}{ds} &= \frac{\partial H_\textrm{QD}}{\partial \mb{p}_c} = \frac{\mb{p}_c}{m_e^*} -\frac{\omega_c}{2} \hat{\mb{z}} \times \mb{r}_c,   \\
\frac{d \mb{p}_c}{ds} &= - \frac{\partial H_\textrm{QD}}{ \partial \mb{r}_c} =   -\frac{m_e^* \omega_c^2}{4} \mb{r}_c + \frac{\partial}{\partial \mb{r}} U_\textrm{QD}(\mb{r}, s)|_{\mb{r}_c} - \frac{\omega_c}{2}\hat{\mb{z}} \times \mb{p}_c. \\
\end{aligned}
\end{equation}
Therefore, the ground state at the initial time $s=0$ evolves in time, propagating along  
the $\mb{E} \times \mb{B}$ drift determined by $\partial U_\textrm{QD}(\mb{r}, s)/ \partial \mb{r} $ and the magnetic field. The shape (the width and the anisotropy) of the Gaussian wave packet are determined by $A(s)$, $B(s)$ and $C(s)$, which are governed by the differential equations
\begin{equation} \label{eom_c}
\begin{aligned}{}
\frac{d A}{d s} &= \frac{i}{4}  (-1 + 4A^2) \omega_c + \frac{1}{2} (-C + i C^2/2) \omega_c -\frac{i \kappa_+^2(s)}{\omega_c},  \\
\frac{d B}{d s} &= \frac{i}{4}  (-1 + 4 B^2) \omega_c + \frac{1}{2} (C + i C^2/2) \omega_c -\frac{i \kappa_-^2(s)}{\omega_c},  \\
\frac{d C}{d s} &= \omega_c(A -B)  +i\omega_c (A +B)  C,
\end{aligned}
\end{equation}
where $\kappa_{\pm}(s) \equiv [ (\partial_{xx}^2 + \partial_{yy}^2) U_\textrm{QD}|_{\mb{r}_c}/2 \pm \{((\partial_{xx}^2- \partial_{yy}^2) U_\textrm{QD}|_{\mb{r}_c} )^2/4+ {( \partial_{xy}^2  U_\textrm{QD}|_{\mb{r}_c} )^2} \}^{1/2}]^{1/2}/\sqrt{m_e^*}$. 
The state in Eq.~\eqref{init_state} provides the initial condition of the differential equations in Eq.~\eqref{eom_c}. The solution of Eq.~\eqref{eom_c} is
\begin{equation}
\begin{aligned}{} \label{sol}
A(s) &= -\frac{\kappa_{+}(s)}{\kappa_{+}(s)+\kappa_{-}(s)}  -\frac{ \kappa_+(s) (\kappa_+ (s) + \kappa_- (s))}{2\omega_c^2} + \bo(\frac{\kappa_\pm^4}{\omega_c^4}) + \bo(\frac{1}{\kappa_\pm \omega_c}\frac{d\kappa_{\pm}}{ds}), \\
B(s) &= -\frac{\kappa_{-}(s)}{\kappa_{+}(s)+\kappa_{-}(s)} -\frac{\kappa_-(s) (\kappa_+ (s) + \kappa_- (s))}{2\omega_c^2} + \bo(\frac{\kappa_\pm^4}{\omega_c^4}) + \bo(\frac{1}{\kappa_\pm \omega_c}\frac{d\kappa_{\pm}}{ds}), \\
C(s) &= i \frac{\kappa_{+}(s)-\kappa_{-}(s)}{\kappa_{+}(s)+\kappa_{-}(s)} +i \frac{(\kappa_{+}(s)-\kappa_{-}(s))(\kappa_+ (s) + \kappa_- (s))}{4 \omega_c^2}  + \bo(\frac{\kappa_\pm^4}{\omega_c^4}) + \bo(\frac{1}{\kappa_\pm \omega_c}\frac{d\kappa_{\pm}}{ds}).
\end{aligned}
\end{equation}
Note that the errors $\bo(\cdots)$ are small under the conditions of $U_\textrm{QD}$ mentioned above; the smoothness of $U_\textrm{QD}$ in the length scale $l_B$, $U_\textrm{QD}(\mb{r} -\mb{r}_0, s) -U_\textrm{QD}(\mb{r}_0 s) \ll \hbar \omega_c$ for $|\mb{r} -\mb{r}_0| <l_B $, implies $\kappa_\pm \ll \omega_c$, while the smoothness of $U_\textrm{QD}$ in the time scale $\omega_c^{-1}$, $\partial U_\textrm{QD}/\partial s \ll \omega_c U_\textrm{QD} $, implies $\partial \kappa_{\pm}/\partial s \ll \omega_c \kappa_\pm$. 
Under the conditions, the second terms of Eqs. \eqref{sol} are also negligible as they become much smaller than the first terms. Then, the solution shows that the time evolved wave packet has an anisotropic Gaussian form of width of $l_B [1 - (\kappa_+-\kappa_-)/(\kappa_+ +\kappa_-)]^{-1/2}$ along the major axis and $l_B [1 +(\kappa_+-\kappa_-)/(\kappa_+ +\kappa_-)]^{-1/2}$ along the minor axis. 
When $U_\textrm{QD}$ is an isotropic harmonic potential, the Gaussian form is also isotropic with $\kappa_+ = \kappa_-$. For general anisotropic QD confinements, the Gaussian form is anisotropic, but this effect is not significant in realistic QDs as discussed below. Thus, $\psi (\mb{r}, s)$ is well approximated as
\begin{equation}
\psi_c(\mb{r}, s) =\frac{1}{\sqrt{\pi l_B^2}} \exp [-\frac{(\mb{r}-\mb{r}_c (s))^2}{2 l_B^2}+i \mb{p}_c (s) \cdot \mb{r}]. \nonumber 
\end{equation}

In order to elucidate the behavior and the validity of the result, we discuss the time evolution of the ground state in a time-dependent anisotropic harmonic QD, $U_\textrm{QD}(\mb{r}, s) = (\omega_{0,x}^2(s) x^2 +\omega_{0,y}^2(s) y^2)/(2 m_e^*) + F(s) x$.
Here, the force $F(s)$ describes a time-dependent shift of the center of the confinement in the $x$ direction. In this case, the error terms $\bo(\cdots)$ in Eq.~\eqref{sol} vanish and the time evolution $\psi(\mb{r}, s)$ is obtained from Eqs.~\eqref{finite_time_evol}, \eqref{eom_rp}, and \eqref{sol} as 
\begin{equation}
\begin{aligned}{}
\psi (\mb{r}, s) &= \exp \Big[ -\Big\{\frac{(x-x_c(s))^2}{ l_B^2}\frac{\omega_{0,x}(s)}{\omega_{0,x}(s) +\omega_{0, y}(s)}+\frac{(y-y_c(s))^2}{ l_B^2}\frac{\omega_{0,y}(s)}{\omega_{0,x}(s) +\omega_{0, y}(s)}\Big\} \big(1+ \frac{(\omega_{0,x}(s) +\omega_{0, y}(s))^2}{2\omega_c^2} \big)  \\
&\quad\quad \quad +i \frac{(x-x_c(s)) (y-y_c(s))}{l_B^2} \frac{\omega_{0,x}(s)-\omega_{0,y}(s)}{\omega_{0,x}(s) +\omega_{0, y}(s)} \big(1+\frac{(\omega_{0,x}(s) +\omega_{0, y}(s))^2}{ 4 \omega_c^2 }\big)\Big] e^{i \mb{p}_c(s) \cdot \mb{r}}.
\end{aligned} \label{EXACT}
\end{equation}
We note that when $\omega_{0,x(y)}(s)$ is time independent, Eq.~\eqref{EXACT} is identical to the result analytically obtained~\cite{Madhav-2} by diagonalizing the Hamiltonian.
The overlap between $\psi$ and $\psi_c$ is obtained as
\begin{equation} \label{ovl}
|\ovl{\psi_c(s)}{\psi(s)}|^2 \approx 1 - \left(\frac{\omega_{0,x}(s) +\omega_{0,y}(s)}{2 \omega_c}\right)^4 -\frac{1}{2} \left(\frac{\omega_{0,x}(s)-\omega_{0,y}(s)}{\omega_{0,x}(s) +\omega_{0,y} (s)} \right)^2.
\end{equation}
The second term means that $\psi$ is well approximated by the Gaussian packet Eq.~\eqref{finite_time_evol} in a sufficiently strong magnetic field, and the last term shows that $\psi(s)$ is well described by the isotropic Gaussian packet $\psi_c$ when $U_\textrm{QD}$ is not too anisotropic.
For example when $|\omega_{0,x}(s)-\omega_{0,y}(s)|$ is $\sim$30\% of $(\omega_{0,x}(s)+\omega_{0,y}(s))/2$ (this is the value that we find in the numerical simulation of an realistic QD pump in Fig.~1), the second term of Eq.~\eqref{ovl} is less than $10^{-2}$. 

}

\section{Time dependent scattering theory}

We here describe the scattering theory for the QD pump, which follows a Floquet theory~\cite{floquet-2},
and derive the emitted wave functions in Eqs.~(3) and (5). 

\begin{figure}[t]
\includegraphics[width=0.7\textwidth]{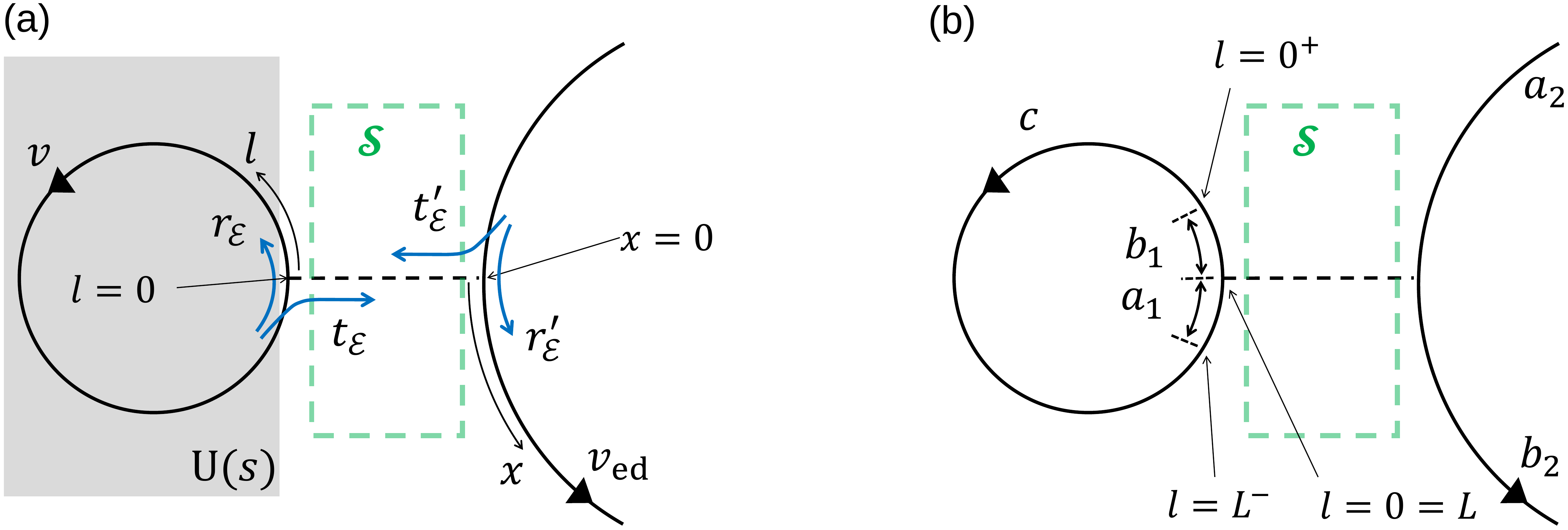}
\caption{Scattering problem (a) before and (b) after the gauge transformation where the information of the time dependence of the potential $U$ is attached onto the scattering amplitudes.}
\label{figs1}
\end{figure}

In the regime of $v_U \tau_d \gtrsim \Delta$, the QD pump is simplified into a scattering model in Fig.~\ref{figs1}.
The coherent state propagates along a loop of coordinate $l$ which couples with the edge of the 2DEG outside the QD via the exit barrier located at $l=0$.
The loop represents the trajectory of the state at $s > s_\textrm{mx}$. 
The state gains the potential energy $U(s)$ by $V_\textrm{G1}$. 
$U(s)$ increases linearly and then stays at its maximum $U_\textrm{mx}$ at $s > s_\textrm{mx}$, following $V_\textrm{G1}(s)$ in Fig.~1.
The exit barrier is parameterized by scattering amplitudes $t_\mathcal{E}$, $r_\mathcal{E}$, $t'_\mathcal{E}$, and $r'_\mathcal{E}$ connecting plane waves of the QD ($\sim Y e^{i k l}$, $Y$ being the wave function in the transverse direction) and those of the 2DEG edge ($\sim Y e^{i k x}$) at the same energy $\mathcal{E}$.
The 2DEG edge states belong to the lowest Landau level, since the edge state of the higher levels are located farther from the exit barrier (by distance longer than $l_B$) hence the coupling from the QD to them is much weaker.
At time $s_0$, the coherent state has so low energy that $t_\mathcal{E} = 0$ within its energy window, and is located at $l=0^{-}$ for simplicity.

We solve the scattering problem, using the gauge transformation where the dynamical phase by $U(s)$ is attached onto the scattering amplitudes, as in a Floquet theory~\cite{floquet-2}.
We consider a phase $\Lambda (l,s) = \Theta(l-0^{+}) \Theta(L^{-}-l) \int_{-\infty}^s U(s')ds' $,
where $L$ is the total length of the loop, $\Theta (x) = 1$ for $x > 0$, and $\Theta (x) = 0$ for $x < 0$; in this Supplementary Materials, we use the convention of $\hbar \equiv 1$. Then the potential $U$ and the vector potential $\mb{A}$ are gauge transformed as
\begin{equation}
\begin{aligned}{}
\Phi =U(t) \,\,\,\,\,\, &\rightarrow \,\,\,\,\,\, \Phi - \partial \Lambda/\partial s =0 \\
\mb{A} =0 \,\,\,\,\,\, &\rightarrow \,\,\,\,\,\, \mb{A} + \nabla \Lambda = [\delta(l-0^{+})-\delta(l-L^{-})] \int_{-\infty}^s U(s')ds'.
\end{aligned}
\end{equation}
After the transformation, the loop becomes time independent, and instead the coupling at $l=0$ (at $l=0^+$ and $l=L^-$) between the loop and the 2DEG edge becomes time dependent, carrying the information of $U(s)$. 

Then, to apply the gauge transformation, we decompose the loop into three regions, $l \in [L^-, L]$, $l \in [ 0, 0^+]$, and $l \in [0^+, L^-]$,  and we assign state amplitudes $a_1$, $b_1$, and $c$, to the regions; cf. Fig.~\ref{figs1}(b). For example, a scattering state incoming from the 2DEG edge can be decomposed into an incident edge state of amplitude $a_{2, \mathcal{E}}$, an edge state with amplitude $b_{2, \mathcal{E}}$ outgoing from the coupling point, a loop state with amplitude $b_{1, \mathcal{E}}$ in $l \in [ 0, 0^+]$, a loop state with amplitude $a_{1, \mathcal{E}}$ in $l \in [L^-, L]$, and a loop state with amplitude $c_\mathcal{E}$ in $l \in [0^+, L^-]$. Here, $\mathcal{E}$ is the energy of the incident state. At $l=0^+$ and $l=L^-$, the information of the time-dependent $U(s)$ is attached to wave functions such that a wave function $\psi_{c, \mathcal{E}}(l,s)$ of energy $\mathcal{E}$ in $l \in [0^+, L^-]$ couples with $\psi_{b_1, \mathcal{E}}(s)$ in $l \in [0,0^+]$ and $\psi_{a_1, \mathcal{E}}(s)$ in $l \in [L^-,L]$ as 
\begin{align}
\psi_{c, \mathcal{E}}(l=0^+, s) &= \psi_{b_1,\mathcal{E}} (s)  e^{i\phi(s)}, \label{cb1} \\
\psi_{a_1, \mathcal{E}} (s) &= \psi_{c, \mathcal{E}} (l=L^-, s) e^{-i\phi(s)}, \nonumber
\end{align}
where $\phi (s) = \int_{-\infty}^s U(s') ds'$. Since $\psi_{c, \mathcal{E}}(l = L^-,s) = \psi_{c, \mathcal{E}} (l = 0^+, s - \tau_d)$, one finds
\begin{equation}
\psi_{a_1, \mathcal{E}} (s) = e^{-i\phi(s) +i\phi(s-\tau_d)}  \psi_{b_1, \mathcal{E}} (s-\tau_d). \nonumber \end{equation}
From the Fourier transformation of this, the relation between the amplitudes $a_1$ and $b_1$ is found as
\begin{align}
a_{1,\mathcal{E}} (E') &= \int dE'' g(E'-E'') b_{1, \mathcal{E}}(E'')  e^{i E'' \tau_d} \label{a1}
\end{align}
where $g(E) \equiv \int dt ^{-i\phi(t)+i\phi(t-\tau_d)} e^{iE t}$.
And, $a_1$ and $a_2$ are related with $b_1$ and $b_2$ as
\begin{align}
b_{1, \mathcal{E}}(E') &=  \delta (E'-\mathcal{E}) t'_{\mathcal{E}} a_{2,\mathcal{E}} + e^{i\phi_\textrm{AB}} r_{E'} a_{1, \mathcal{E}}(E') \label{S1} \\
b_{2, \mathcal{E}} (E') &= \delta(E'-\mathcal{E})  r'_{\mathcal{E}} a_{2, \mathcal{E}} + t_{E'} a_{1 \mathcal{E}}(E').  \nonumber
\end{align}
Note that the Aharonov-Bohm phase $\phi_\textrm{AB} = 2\pi B \pi (l/2\pi)^2/(h/e) $ is attached to the reflection event in the loop.


Next, we derive the Fabry-Perot type scattering state resulting from the incident state.
Combining Eqs.~\eqref{a1} and \eqref{S1}, we find the recursive equations for $b_1$ and $b_2$,
$b_{1, \mathcal{\mathcal{E}}}(E') = \delta (E'-\mathcal{E}) t'_{\mathcal{E}} a_{2,\mathcal{E}} + e^{i\phi_\textrm{AB}} r_{E'} \int dE'' g(E'-E'') e^{i E'' \tau_d}  b_{1, \mathcal{E}}(E'')$,
$ b_{2,\mathcal{E}}(E')=   \delta(E'-\mathcal{E}) r'_\mathcal{E} a_{2, \mathcal{E}} +t_{E'} \int dE'' g(E'-E'') e^{iE'' \tau_d} b_{1, \mathcal{E}}(E'')$.
Their Fourier transformations are
\begin{align}
b_{1, \mathcal{E}}(s) &=  t'_\mathcal{E} e^{-i \mathcal{E} s} a_{2, \mathcal{E}} + e^{i\phi_\textrm{AB}} \int ds' r(s') e^{-i \{\phi(s-s') -\phi(s-s'-\tau_d) \}} b_{1, \mathcal{E}}(s-s'-\tau_d ) \label{b1e22} \\ 
b_{2, \mathcal{E}}(s) &=   r'_\mathcal{E} e^{-i \mathcal{E} s} a_{2, \mathcal{E}} + \int ds' t(s') e^{-i \{\phi(s-s') -\phi(s-s'-\tau_d) \}} b_{1, \mathcal{E}}(s-s'-\tau_d ). \nonumber
\end{align}
$r(s) \equiv \int dE r_E e^{-i Es}$ and $t(s) \equiv \int dE t_E e^{-i Es}$ are the Fourier transforms of $r_E$ and $t_E$, and it can be approximated as a peak structure with width $1/\Delta$ at $s=0$.
The integral in Eq.~\eqref{b1e22} is further evaluated by using the peak structure and under 
the condition of $\ddot{U} / (2\dot{U}) \ll \Delta$,
$\int ds' r(s') e^{-i \{\phi(s-s')-\phi(s-s'-\tau_d)\}} e^{i \mathcal{E} s'} = \int ds' r(s') \exp[-i \{ \phi(s)-\phi(s-\tau_d) - (U(s) -U(s-\tau_d))s' +\frac{\dot{U}(s)-\dot{U}(s-\tau_d)}{2} (s')^2 + \bo(s'^3) \} ]  e^{i \mathcal{E} s'}\approx r_{\mathcal{E} +U(s)-U(s-\tau_d)} e^{-i \{\phi(s)-\phi(s-\tau_d)\}} $; in the first equality, we use the Taylor expansion of $\phi(s-s')$ and $\phi(s-s'-\tau_d)$ at $s'=0$, considering the peak structure of $r(s')$ at $s'=0$; in the second equality, we ignore the quadratic term, applying the condition of $\ddot{U} / (2\dot{U}) \ll \Delta$. Then, $b_{1, \mathcal{E}}$ in Eq.~\eqref{b1e22} is iteratively solved as
\begin{equation} \label{b1}
b_{1, \mathcal{E}}(s) = t'_\mathcal{E} e^{-i\mathcal{E} s} a_{2, \mathcal{E}} +\sum_{M=1}^{\infty} e^{-i\phi(s)+i\phi(s-M\tau_d)} e^{i M \mathcal{E}\tau_d} e^{iM \phi_\textrm{AB}} \left[ \Pi_{m'=1}^M  r_{\mathcal{E}+ U(s-(M-m'-1)\tau_d) -U(s-M\tau_d)} \right] t'_\mathcal{E} e^{-i\mathcal{E} s} a_{2, \mathcal{E}} 
\end{equation}
Note that the condition of $\ddot{U} / (2\dot{U}) \ll \Delta$ is satisfied in usual experiments, because the smallest time scale for variation of $U$ is limited by the band width of signal generator. Namely, the smallest time scale is $\sim 50$ ps for 10 GHz bandwidth, and then $\ddot{U} / (2\dot{U}) \sim 0.02$ meV, while $\Delta \sim 0.5$ meV (see the main text). 
Plugging Eq.~\eqref{b1} into Eqs.~\eqref{cb1} and \eqref{b1e22}, we obtain the Fabry-Perot type expression of $c$ and $b_2$ in time domain, 
\begin{align}
&c_\mathcal{E} ( s) = e^{i\phi(s)}  t'_\mathcal{E} e^{-i\mathcal{E} s} a_{2, \mathcal{E}} +\sum_{M=1}^{\infty} e^{i\phi(s-M\tau_d)} e^{i M \mathcal{E}\tau_d} e^{iM \phi_\textrm{AB}} \left[ \Pi_{m'=1}^M  r_{\mathcal{E}+ U(s-(M-m'-1)\tau_d) -U(s-M\tau_d)}   \right]  t'_\mathcal{E} e^{-i\mathcal{E} s} a_{2, \mathcal{E}} \label{ct}  \\
&b_{2, \mathcal{E}}(s) =  r'_{\mathcal{E}} e^{-i\mathcal{E} s}a_{2, \mathcal{E}}  \nonumber \\
&+ \sum_{M=1}^{\infty} t_{\mathcal{E}+ U(s)-U(s-M\tau_d) }  e^{-i\phi(s)+i\phi(s-M\tau_d)}  
  e^{iM \mathcal{E} \tau_d} e^{i (M-1) \phi_\textrm{AB}} [ \Pi_{m'=1}^{M-1}   r_{\mathcal{E}+ U(s-(M-m'-1)\tau_d) -U(s-M\tau_d)}  ] t'_\mathcal{E} e^{-i\mathcal{E} s} a_{2, \mathcal{E}}  \label{b2t}
\end{align}
Here each term of index $M$ in Eq.~\eqref{b2t} describes the process that the incident electron with amplitude $a_{2,\mathcal{E}}$ enters the loop at time $s-M\tau_d$, circles the loop $M$ times, and then escapes from the loop at time $s$.  In the $M=1$ term we use $\prod_{m'=1}^{0}\equiv 1$ instead of 0, for brevity. 

Next, we determine the incident amplitudes $a_{2,\mathcal{E}}$ 
with which the resulting scattering state is identical to the coherent state $\psi_\textrm{coh} =(\sqrt{\pi/2} v\tau_p)^{-1/2} \exp[-l^2/(v\tau_p)^2] \exp[i\epsilon_0 l /v ]$ at time $s_0$,
$\int d \mathcal{E} \psi_{c, \mathcal{E}}(l, s_0) = \psi_\textrm{coh}(l)$. We compute $a_{2,\mathcal{E}}$, using  $\psi_{c, \mathcal{E}}(l, s_0) = c_{\mathcal{E}}(s_0-l/v)$ and Eq.~\eqref{ct}, and choosing the time dependence of $U(s)$ at $s \le s_0$ as $U(s\le s_0) = U(s_0)$ to simplify the calculation (note that the emitted wave function of $b_{2, \mathcal{E}}(s)$ at $s_\textrm{rd} \ge s_0$ does not rely on this specific choice):
\begin{eqnarray} 
a_{2,\mathcal{E}} & \propto & e^{-i\phi(s_0)} e^{i \mathcal{E} s_0 }   e^{-(\mathcal{E}-U(s_0)-\epsilon_0)^2 \tau_p^2/4}  \frac{(t'_\mathcal{E})^*}{1-r_{\mathcal{E}} e^{-i(\mathcal{E}-U(s_0))\tau_d -i\phi_\textrm{AB}}} \label{coeff} \\
& = & e^{-i\phi(s_0)} e^{i \mathcal{E} s_0 }   e^{-(\mathcal{E}-U(s_0)-\epsilon_0)^2 \tau_p^2/4} \sum_n \frac{(t'_\mathcal{E})^*}{|t_\mathcal{E}|^2/2 - i (\mathcal{E} - \mathcal{E}_n) \tau_d },
\nonumber \end{eqnarray}
where $\mathcal{E}_n = 2 \pi \hbar n / \tau_d + U(s_0) -\phi_\textrm{AB}\hbar/\tau_d$ is the resonance energy of the loop at $s_0$.
In the second equality, the expression is decomposed into the resonance states, since $r_\mathcal{E} \to 1$ at $s_0$. 

Finally, we compute the emitted state of $\psi(x,s) = \int d\mathcal{E}  b_{2, \mathcal{E}}(s_\mathrm{rd})$, where the $s_\mathrm{rd} \equiv s - x/v_\mathrm{ed}$, and
obtain Eq.~(3)
\begin{equation} \label{gew}
\begin{aligned}{}
\psi(x,s) &\propto \sum_{n,m=1}^{\infty}
 e^{-(\mathcal{E}_n-\epsilon_0)^2 \tau_p^2/4}
 t_{\mathcal{E}_n+U(s_\mathrm{rd}) -U(s_0) } e^{i m\phi_\textrm{AB}} \left[ \Pi_{m'=0}^{m-1} r_{\mathcal{E}_n+U(s_\mathrm{rd}-(m-m') \tau_d) -U(s_0) } \right] \\
& \quad \quad \quad \times e^{-i (\mathcal{E}_n -U(s_0)) (s_\mathrm{rd}-s_0)} e^{im(\mathcal{E}_n-U(s_0))\tau_d  } e^{-i \phi(s_\mathrm{rd}) + i \phi (s_0)} \zeta_m (s_\mathrm{rd}),
\end{aligned}
\end{equation}
where  $\zeta_{m}(s_\mathrm{rd}) = 1$  for $s_\mathrm{rd} \in [s_0+ m \tau_d, s_0 + (m+1) \tau_d )$ and $\zeta_m (s_\mathrm{rd}) = 0$ otherwise. Here, we performed the integral $\int d\mathcal{E}   b_{2, \mathcal{E}}(s_\mathrm{rd})$, based on the fact that the Lorentzian function in the integrand becomes the delta function when $r_\mathcal{E} \to 1$. 
Note that Eq.~\eqref{gew} does not include the emission process with no circulation, since we choose $s_0$ such that the energy $\epsilon_0$ of the coherent state is much lower than the exit barrier height and hence $t_{\epsilon_0} \ll 1$.

Next, we derive Eq.~(5) when $\tau_\mr{tun} \gg \tau_p$. This limit is achieved when $|t_{\epsilon_0 + U_\textrm{mx}}|$ is small. 
Then the emission occurs during the time of $U(s) = U_\textrm{mx}$, as $|t_\mathcal{E}|$ is more smaller at $s < s_\textrm{mx}$. 
This allows us to write the product term in Eq.~\eqref{gew} as an exponential decaying function in $s_\mathrm{rd}$ as 
$\Pi_{m'= 1}^{m} r_{\mathcal{E}_n+U(s_\mathrm{rd}-  m')\tau_d) -U(s_0)} \approx (r_{\mathcal{E}_n+U(s_\mathrm{mx}) -U(s_0)})^{\lfloor (s-s_\mathrm{mx})/\tau_d \rfloor} 
\approx \exp[\, (\ln r_{\mathcal{E}_n+U(s_\mathrm{mx}) -U(s_0)}) \,(s_\mathrm{rd}-s_\mathrm{mx} )/\tau_d \, ]
 \approx    \exp[\,- (s_\mathrm{rd}-s_\mathrm{mx})  |t_{\mathcal{E}_n+U(s_\mathrm{mx}) -U(s_0)}|^2 /(2\tau_d) \,]$, where $\lfloor x \rfloor$ means the integer obtained by flooring x; in the first approximation, the reflection amplitudes before $s_\textrm{mx}$ are replaced by 1 and in the other approximations, it is used that $r_{\mathcal{E}_n+U(s_\textrm{mx}) -U(s_0)}$ is close to 1. And, the phase gain by $U$ reduces to $e^{-i\phi (s_\mathrm{rd})+i\phi(s_0)} \propto e^{-i U(s_\mathrm{mx}) s_\mathrm{rd}  }$
Then, the Eq. \eqref{gew} is expressed as
\begin{equation}\label{ew_res}
\psi(x,s) \propto 
\sum_{n=1}^{\infty} e^{-(\mathcal{E}_n-\epsilon_0)^2 \tau_p^2/4}   t_{\mathcal{E}_n  +U(s_\mathrm{mx})-U(s_0)} e^{-\frac{ |t_{\mathcal{E}_n+U(s_\mathrm{mx})-U(s_0)}|^2}{2\tau_d} (s_\mathrm{rd}-s_\mathrm{mx} )}  e^{-i (\mathcal{E}_n+U(s_\mathrm{mx})-U(s_0)) s_\mathrm{rd}} \Theta(s_\mathrm{rd}-s_\mathrm{mx}),
\end{equation}
where $\Theta(x)$ is 1 for $x>0$ and 0 otherwise. 
Each $n$ term of this equation describes the emission of the $n$-th resonance state whose life time is $\tau_d/|t_{\mathcal{E}_n +U(s_\mathrm{mx})-U(s_0)}|^2$. 
The equation~(5) is derived from Eq.~\eqref{ew_res} using the fact that $t_{\mathcal{E}_n +U(s_\mathrm{mx})-U(s_0)}$ is a rapidly increasing function of $n$ in the limit of $\tau_{\mr{tun}} \gg \tau_p$. 
The detailed steps for the derivation are as follows. 
We first express the transmission amplitude as $t_{\mathcal{E}_n+U(s_\mathrm{mx})-U(s_0)} = t_{\epsilon_0+U(s_\mathrm{mx})-U(s_0)} \exp[\, \tau_\mr{tun} (\mathcal{E}_n -\epsilon_0) \,]$, using the definition of $\tau_\textrm{tun}$.
Then the factor depending on $t_{\mathcal{E}_n+U(s_\mathrm{mx})-U(s_0)}$ in Eq.~\eqref{ew_res}, 
$t_{\mathcal{E}_n+U(s_\mathrm{mx})-U(s_0)}
  \exp[\, - |t_{\mathcal{E}_n+U(s_\mathrm{mx})-U(s_0)}|^2 (s_\mathrm{rd}-s_\mathrm{mx})/(2\tau_d) \, ]$, 
has a non-vanishing value only for the resonance levels $\mathcal{E}_n$
$\in [\epsilon_0  +
( \ln \frac{2\tau_d}{|t_{\epsilon_0+U(s_\mathrm{mx})-U(s_0)}|^2 (s_\mathrm{rd}-s_\mathrm{mx})} )/\tau_\mr{tun}
 - 1/\tau_{\mr{tun}}, \epsilon_0  +
( \ln \frac{2\tau_d}{|t_{\epsilon_0+U(s_\mathrm{mx})-U(s_0)}|^2 (s_\mathrm{rd}-s_\mathrm{mx})} )/\tau_\mr{tun}
 + 1/\tau_{\mr{tun}}]$.
This means that at retarded time $s_\textrm{rd}$, the emitted wave function is determined only by the resonance levels whose lifetime is similar to $s_\textrm{rd}-s_\textrm{mx}$.
When $\tau_p \ll \tau_{\mr{tun}}$, $t_{\mathcal{E}_n+U(s_\mathrm{mx})-U(s_0)}
  \exp[\, - |t_{\mathcal{E}_n+U(s_\mathrm{mx})-U(s_0)}|^2 (s_\mathrm{rd}-s_\mathrm{mx})/(2\tau_d) \, ]$ has a sharp peak of width $1/\tau_\textrm{tun} \ll 1/\tau_p$, thus the gaussian factor in Eq. \eqref{ew_res} can be approximated as  
$e^{-(\mathcal{E}_n-\epsilon_0)^2 \tau_p^2/4}
\approx \exp[\, -\frac{\tau_p^2}{4\tau_\mr{tun}^2}(\ln \frac{|t_{\epsilon_0+U(s_\mathrm{mx})-U(s_0)}|^2 (s_\mathrm{rd}-s_\mathrm{mx})}{2\tau_d})^2 \,] $.
Finally, using the identities,
$\int d\mathcal{E}\,  e^{\mathcal{E}/2} e^{-a e^\mathcal{E}} e^{i \mathcal{E} s} = a^{- 1/2 -i s} \Gamma[ 1/2+ i s] $,
$\Gamma \left[ 1/2+ i s \right] \approx \sqrt{\pi \mr{sech} (\pi s )} \approx \sqrt{\pi/2} \mr{sech}(s/2)$, and
$ \int ds\, \mr{sech}(as) e^{-i \omega s}= \pi \mr{sech} (\frac{\pi}{2a}\omega)/a $,
we obtain Eq.~(5),
\begin{align} \label{ll}
\psi(x, s)   \propto &
\sqrt{\frac{2\tau_d}{(s_\mathrm{rd}-s_\mathrm{mx})|t_{\epsilon_0}|^2}}
e^{-   \left( \frac{\tau_p}{\tau_\mr{tun}} \ln \frac{|t_{\epsilon_0}|^2 (s_\mathrm{rd}-s_\mathrm{mx})}{\tau_d}\right)^2} \nonumber \\ 
&  \times \sum_{n=1}^{\infty} e^{-i (\mathcal{E}_{n} +U(s_\textrm{mx})-U(s_0)) (s_\mathrm{rd}-s_\mathrm{mx})} \mr{sech} \left(\pi \ln  \frac{(s_\mathrm{rd}-s_\mathrm{mx}) |t_{\mathcal{E}_{n} +U(s_\mathrm{mx})-U(s_0)}|^2}{2\tau_d} \right)  \Theta(s_\mathrm{rd}-s_\mathrm{mx}).
\end{align}


\begin{thebibliography}{9}
\bibitem{NOTE1} For the setup in Fig.~\ref{fig1}(a), our numerical simulation shows that when the entrance barrier height reaches the Fermi level, the QD confinement energy is $\hbar \omega_0 \sim 1.3$ meV, the level spacing is $\delta E = \hbar \omega_0^2 / \omega_c \sim 0.07$ meV, and $\omega_c = 24$ meV at 14 T. Then thermal
excitations are negligible at 100 mK (accessible in experiments), as $e^{- \delta E/(k_B T)} \simeq 1.5 \times 10^{-4}$.
\bibitem{NOTE2} According to our numerical simulation, the electron in the QD feels about 10 \% of the strength of $V_{G1}$ and $V_{G2}$, while hot electrons moving along the 2DEG edge outside the QD feel about 50 \% of the strength of $V_{G3}$.
\end{thebibliography}
\end{document}